\documentclass[nofootinbib,twocolumn,aps,prd,amsmath,superscriptaddress]{revtex4-1}
\usepackage{setspace}
\usepackage{color}
\usepackage{fancyhdr}
\usepackage{graphicx}
\usepackage[ansinew]{inputenc}
\usepackage{amssymb}
\usepackage{amsmath}
\usepackage{cancel}
\usepackage{float}
\usepackage{graphicx}
\usepackage{float}
\usepackage{todonotes}
\usepackage{subfigure}
\usepackage{cancel}
\usepackage{tabularx}

\newcommand{\rom}[1]{\MakeUppercase{\romannumeral #1}}
\begin{document}
\title{Classifications of Twin Star Solutions for a Constant Speed of Sound Parameterized Equation of State}
\author{Jan-Erik Christian}
\email{christian@astro.uni-frankfurt.de}
\affiliation{Institut f\"ur Theoretische Physik, Goethe Universit\"at Frankfurt, 
	Max von Laue Strasse 1, D-60438 Frankfurt, Germany}

\author{Andreas Zacchi}
\email{zacchi@astro.uni-frankfurt.de}
\affiliation{Institut f\"ur Theoretische Physik, Goethe Universit\"at Frankfurt, 
Max von Laue Strasse 1, D-60438 Frankfurt, Germany}

\author{J\"urgen Schaffner-Bielich}
\email{schaffner@astro.uni-frankfurt.de}
\affiliation{Institut f\"ur Theoretische Physik, Goethe Universit\"at Frankfurt, 
Max von Laue Strasse 1, D-60438 Frankfurt, Germany}
\date{\today}

  \begin{abstract}
  We explore the possible mass radius relation of compact stars for the equation of states with a first order phase transition. The low density matter is described by a nuclear matter equation of state resulting from fits to nuclear properties. A constant speed of sound parametrization is used to describe the high density matter phase with the speed of sound $c_s^2=1$.
  A classification scheme of four distinct categories including twin star solutions, i. e. solutions with the same mass but differing radii, is found which are compatible with the $M \ge 2M_\odot$ pulsar mass constraint. 
  We show the dependence of the mass and radius differences on the transition parameters and delineate that higher twin star masses are more likely to be accompanied by large radius differences. These massive twin stars are generated by high values of the discontinuity in the energy density and the lowest possible values of the transition pressure that still result in masses of $M \geq 2M_\odot$ at the maximum of the hadronic branch.

  \end{abstract}
\maketitle
  \section{Introduction}
The recently discovered pulsars PSR J1614-2230 \cite{Demorest:2010bx,Fonseca:2016tux} 
and  PSR J0348+0432 \cite{Antoniadis:2013pzd} of 2$M_{\odot}$ have revived 
the discussion on the interior composition of compact stars. 
The equation of state (EoS) of nuclear matter is well understood up to and around nuclear saturation 
density \cite{Akmal:1998cf,Typel:2009sy,Gandolfi:2011xu,Alford:2014dva}.
High density matter appearing in compact stars and the possible 
role of exotic states have been investigated for years in several works on the 
subject \cite{Bodmer:1971we,Witten:1984rs,Weber:2004kj,Lattimer:2006xb,Zacchi:2015lwa}, and the  
corresponding EoSs have to be constrained respecting the new mass limits.
A supposable scenario for a compact star is that the inner core might be 
composed of a phase of deconfined quarks, whereas the outer shell is made 
of hadronic matter. Such an object is called hybrid star \cite{Alford:2004pf,Coelho:2010fv,Chen:2011my,Masuda:2012kf,Yasutake:2014oxa,Dexheimer:2014pea,Buballa:2014jta,Zacchi:2015oma}.
Depending on the features of the transition between the inner and outer parts of hybrid stars,
a so called twin star configuration might arise, i.e. a third 
family of compact stars appears with alike masses as the second 
family branch of normal neutron or strange quark stars \cite{Gerlach:1968zz,Glendenning:1998ag,Schertler:2000xq,SchaffnerBielich:2002ki,Bhattacharyya:2004fn,Alford:2013aca,Alvarez-Castillo:2014dva,Benic:2014jia,Alford:2015dpa,Blaschke:2015uva,Zacchi:2016tjw}.
With space missions such as NICER (Neutron star Interior Composition ExploreR) \cite{2014SPIE.9144E..20A},  
high-precision X-ray astronomy will be able to offer precise measurements of masses and radii of compact stars  \cite{Watts:2016uzu}. The discovery of two stars with the same masses but different radii could be indeed a signal of the existence of twin stars and implying furthermore the presence of a phase transition in ultra-dense matter.\\ 
In this work we explore various EoSs assuming a Maxwell construction, i.e. a sharp phase 
transition from hadronic matter to quark matter and their 
solutions within the Tolman-Oppenheimer-Volkoff-equations, i.e. the mass-radius relations. We use a density dependent 
nuclear matter EoS (DD2 by Typel et. al. \cite{Typel:2009sy}) for the hadronic outer 
layers and a constant speed of sound parametrization (CSS introduced by Alford et. al. \cite{Alford:2013aca}) 
for the quark matter EoS. The quark matter EoS can be parametrized by the transition pressure $p_{trans}$ or the transition 
energy density $\epsilon_{trans}$ and the corresponding jump in energy 
density $\Delta \epsilon$, assuming a constant value of the speed 
of sound $c_s^2$. This gives the possibility of generating twin 
stars solutions depending on these values. We find that twin stars 
can be classified into four categories and investigate the dependency 
of the properties of the EoS and the corresponding values $p_{trans}$ and $\Delta \epsilon$ on 
the mass and radius differences for twin star solutions.
This paper is organized as follows. In Section \ref{eos} we present the EoS and how to model the phase transition via the constant speed of sound parametrization and show how to generate solutions of the TOV equations, i.e. the mass-radius relations. Next, in Section \ref{results} we show our results for twin stars by varying the different parameters and classify twin star solutions in four different categories. We examine the mass and radius differences and show their dependence on the parameters $p_{trans}$ and $\Delta\epsilon$ of the CSS parametrization. Finally, in Section \ref{conclusion} we compare our results with previous work and present our conclusions.
\newpage
   \section{The Equation of State}\label{eos}
 
  \subsection{Constant speed of sound parametrization}
 For all calculations in this paper a first order phase transition from confined hadronic matter to deconfined quark matter is assumed. For the hadronic matter a density dependent equation of state is used (DD2). This EoS combines a microscopic quantum statistical approach with a generalized relativistic mean field approach and was derived by Typel et. al. \cite{Typel:2009sy}. The quark matter is described by assuming a constant speed of sound (CSS). This parametrization was introduced by Alford et. al. \cite{Alford:2013aca} and reads: 
 \begin{equation}
  \epsilon(p) =
 	\begin{cases} 
 		\epsilon_{DD2}(p)	&  p < p_{trans}\\
 		\epsilon_{DD2}(p_{trans})+\Delta\epsilon + c_{QM}^{-2}(p-p_{trans})	& p > p_{trans}\\
 	\end{cases}
 \end{equation} 
 
\begin{figure}
  	\centering
  	\includegraphics[width=0.4\textwidth]{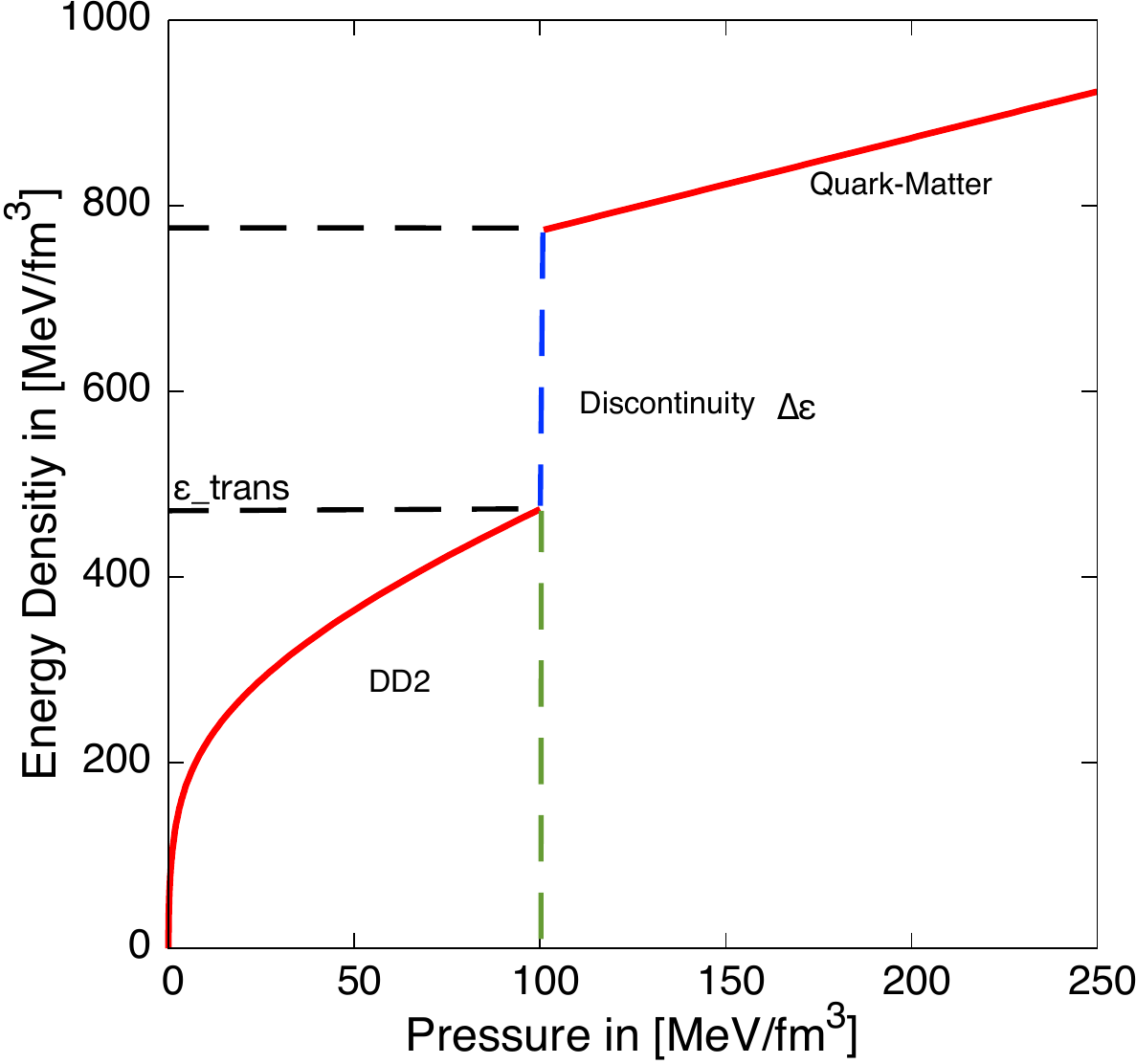}
  	\caption[Equation of state]{\footnotesize The energy density as a function of the transitional pressure is displayed for $\Delta\epsilon$  300 $\rm{MeV/fm^3}$ and $p_{trans}$ 100 $MeV/fm^3$. This configuration provides a third family that can be identified as part of category \rom{3} later in this article.}
  	\label{EoSplot}
  \end{figure}

  Here $\epsilon$ describes the energy density, $p$ the pressure and $p_{trans}$ the pressure at which the transition from the hadronic to quark phase takes place. The transitional pressure $p_{trans}$ has a energy density counterpart $\epsilon_{trans}=\epsilon_{DD2}(p_{trans})$.\\
  The discontinuity at the point of transition is described by $\Delta\epsilon$. This equation of state is displayed in figure \ref{EoSplot}. Setting $c^2_{QM}=1$ there are only two relevant parameters which are $p_{trans}$ and $\Delta\epsilon$.

   \subsection{Hybrid Stars and a Third Family of Compact Stars}
\begin{figure}
  	\centering				
  		\includegraphics[trim = 0mm 10mm 0mm 8mm, clip,width=8.5cm]{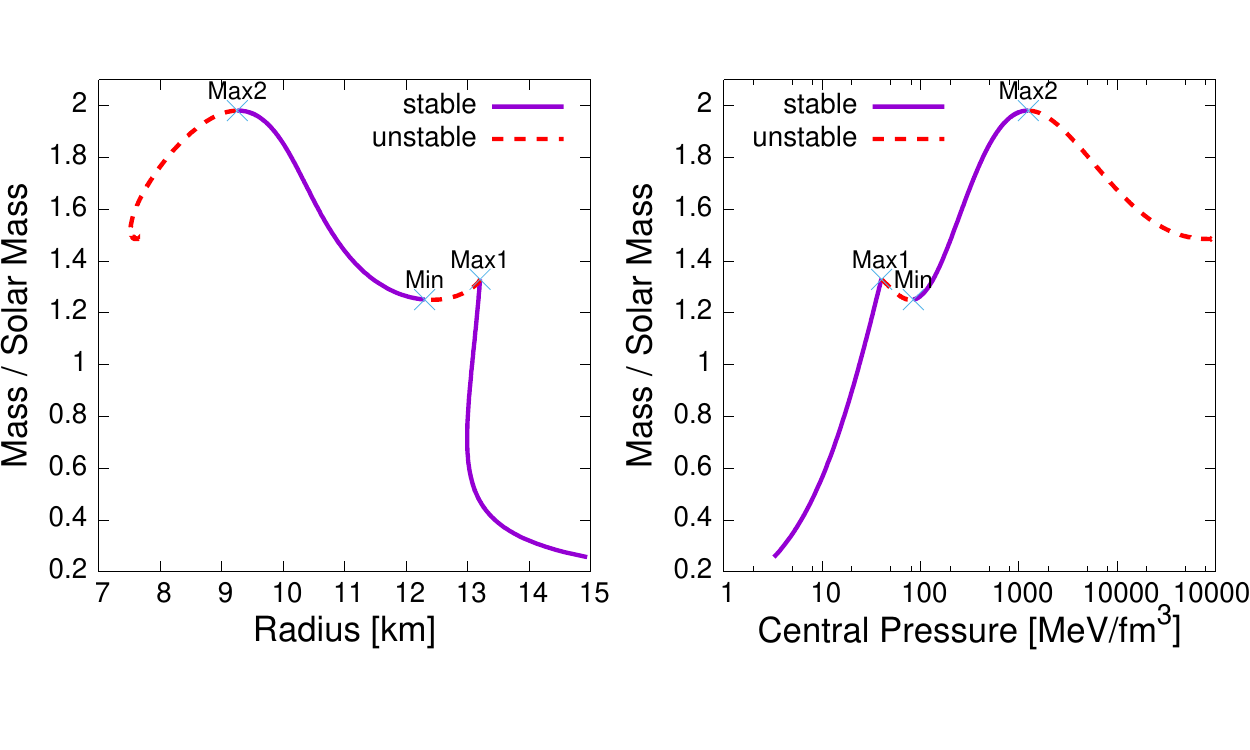}
  		\caption{\footnotesize The left graph is a typical mass-radius relation with the unstable star part of the sequence indicated by the dashed red line. The right graphic depicts the relation between central pressure of a star and its mass. The same parameters where used ($p_{trans}=40\rm{MeV/fm^3}$ and $\Delta\epsilon=368\rm{MeV/fm^3}$).}
  		\label{MinMaxExPlot}
	\end{figure}

 We investigate in the following the effect the variation of the two parameters $p_{trans}$ and $\Delta\epsilon$ on the mass-radius relations and on the stability when they reach central pressures of $p_{central} > p_{trans}$ for $c^2_{QM}=1$.\\
  A good measure for the stability of a sequence of compact stars is the study of the behavior of the mass-radius relation with increasing central pressure. 
  Neutron stars are stable until a maximum in the mass-radius relation is reached. Beyond the maximum the sequence becomes unstable.
  For a detailed review on the stability of compact stars see for example  \cite{1966ApJ...145..505B,Alford:2017vca}.
  In figure \ref{MinMaxExPlot} a representative example of a mass-radius relation and the dependence of the mass with central pressure are provided. Here $\mathrm{Max_1}$ is the maximum of the hadronic branch and $\mathrm{Max_2}$ describes the maximum of the hybrid star branch. The nomenclature for the extrema as depicted there will be used later on in this article.\\
  With this understanding of stability it becomes obvious that at a central pressure equal to $p_{trans}$ one of two things can happen. Either the star becomes unstable or it remains stable even with a quark core. The deciding factor for this distinction is the discontinuity in energy density $\Delta\epsilon$. If  $\Delta\epsilon$ is too low the quark core is not of great influence.
  However, for large jumps in energy density at the transition the star becomes unstable immediately when $p_{central}=p_{trans}$ is reached. 
  \\This condition takes the mathematical form
  \\
  \begin{equation}\label{Seidov-limit}
  \frac{\Delta\epsilon_{crit}}{\epsilon_{trans}}=\frac{1}{2}+\frac{3}{2}\frac{p_{trans}}{\epsilon_{trans}}.
  \end{equation} 
  \\
which is sometimes referred to as Seidov-limit \cite{seidov:1971pty}. $\Delta\epsilon_{crit}$ is the threshold value below which there is a stable hybrid star branch connected to the hadronic star branch \cite{Alford:2014dva, Alford:2015dpa, Alford:2015gna,Zacchi:2015oma}.
$\epsilon_{trans}$ and $p_{trans}$ are the values of the energy density and pressure 
at the phase transition. For a derivation and discussion of 
(\ref{Seidov-limit}) see \cite{Kaempfer:1981a,Kaempfer81,
Kaempfer82,Kaempfer83,Kaempfer83a,zdunikhaenselschaeffer:1983pti,
Lindblom:1998dp}.\\ 
  
  When the Seidov-limit is reached the sequence of stars becomes unstable immediately. However, with the EoSs described above it is possible to regain stability resulting in a second stable sequence of stars if the parameters $p_{trans}$ and $\Delta\epsilon$ are chosen accordingly. This second branch is usually referred to as the "third family" for its property of being the third stable solution of the TOV equations besides white dwarfs and regular neutron stars. In order to be classified as a third family solution the second branch has to be disconnected from the first one. Combinations of $p_{trans}$ and $\Delta\epsilon$, that result in twin star solutions, below the Seidov-limit contain hybrid stars in both branches while sequences above that only contain hybrid stars in the second branch (see figure \ref{allowed1}).\\	
  Third family solutions contain twin stars, these stars are of equal mass with varying radii. Usually one of these twin stars is a regular neutron star located in the first branch whereas the other is always a hybrid star in the second disconnected branch, but is also possible to find a pair of stars with nearly identical mass that are both located in the second branch.\\\\

  \section{Results}\label{results}
  \subsection{Area Containing Twin Star Solutions}
  	\begin{figure}[H]
  		\centering
  		\includegraphics[width=8cm]{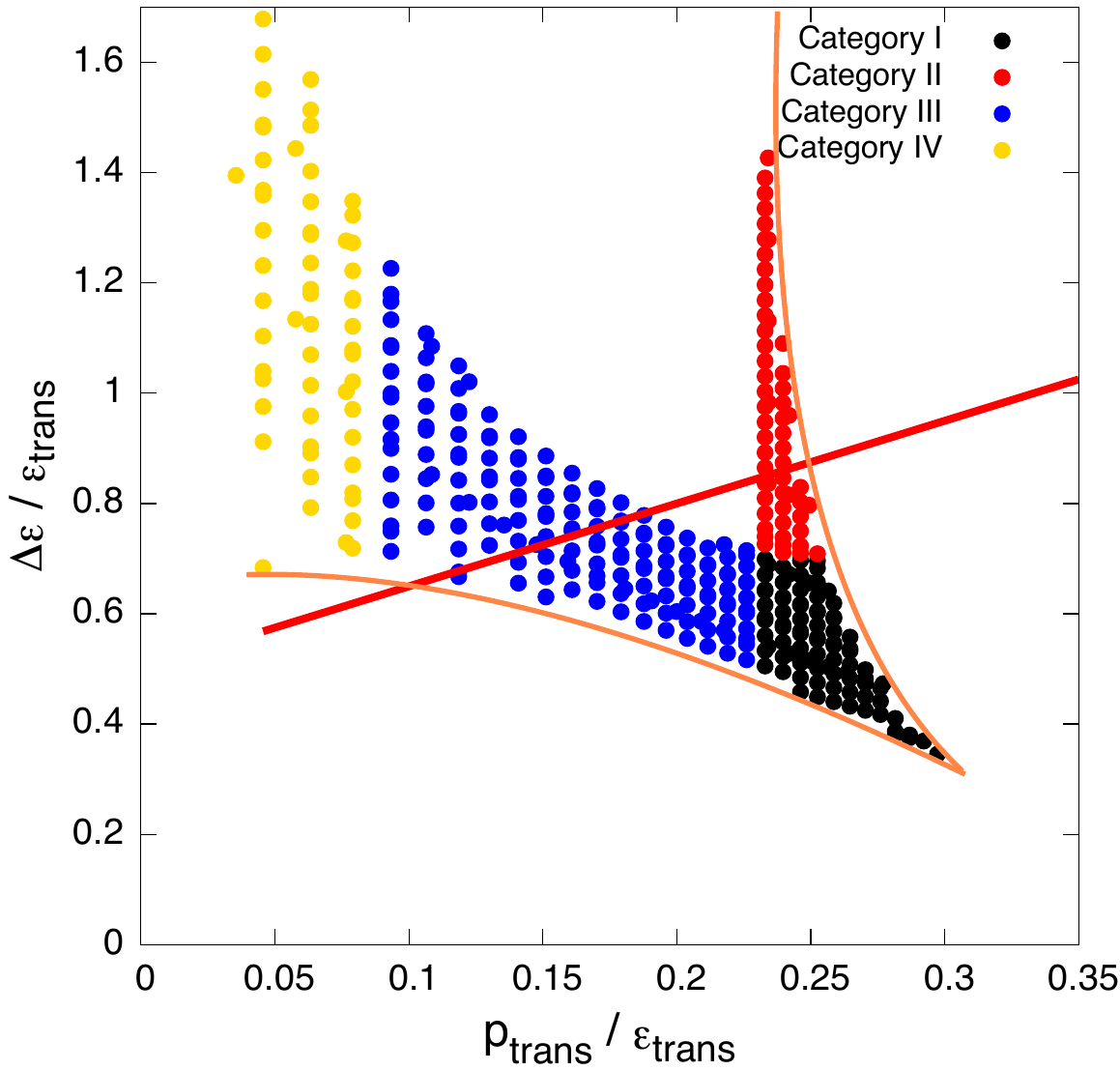}
  		\caption{\footnotesize The parameter area containing twin star solutions is depicted. The points denote all calculated combinations of the parameters $p_{trans}$ and $\Delta\epsilon$ that lead to a third family. Their coloration indicates the categories to be examined later. The red line stands for the Seidov-limit.}
  		\label{allowed1}
  	\end{figure}%
  Only a few combinations of transitional pressure $p_{trans}$ and discontinuity in energy density $\Delta\epsilon$ lead to third family solutions. In figure \ref{allowed1} the combinations of $p_{trans}$ and $\Delta\epsilon$ containing twin star solutions are shown. The red straight line is the Seidov-limit \eqref{Seidov-limit}. The plus signs are mass-radius relations generated by distinct $p_{trans}$ and $\Delta\epsilon$ that contain an additional stable branch. We see that most third family solutions are above the Seidov-limit. I.e. most mass-radius relations with twin stars do not contain hybrid stars in their first branch. Alford et. al. \cite{Alford:2013aca} find a very similar area for $c_s^2=1$. With $c_s^2=\frac{1}{3}$ a much smaller parameter space would generate third family solutions. 	
  \subsection{Classification by Mass}
\begin{figure}
  	\centering
  	\includegraphics[width=8.5cm]{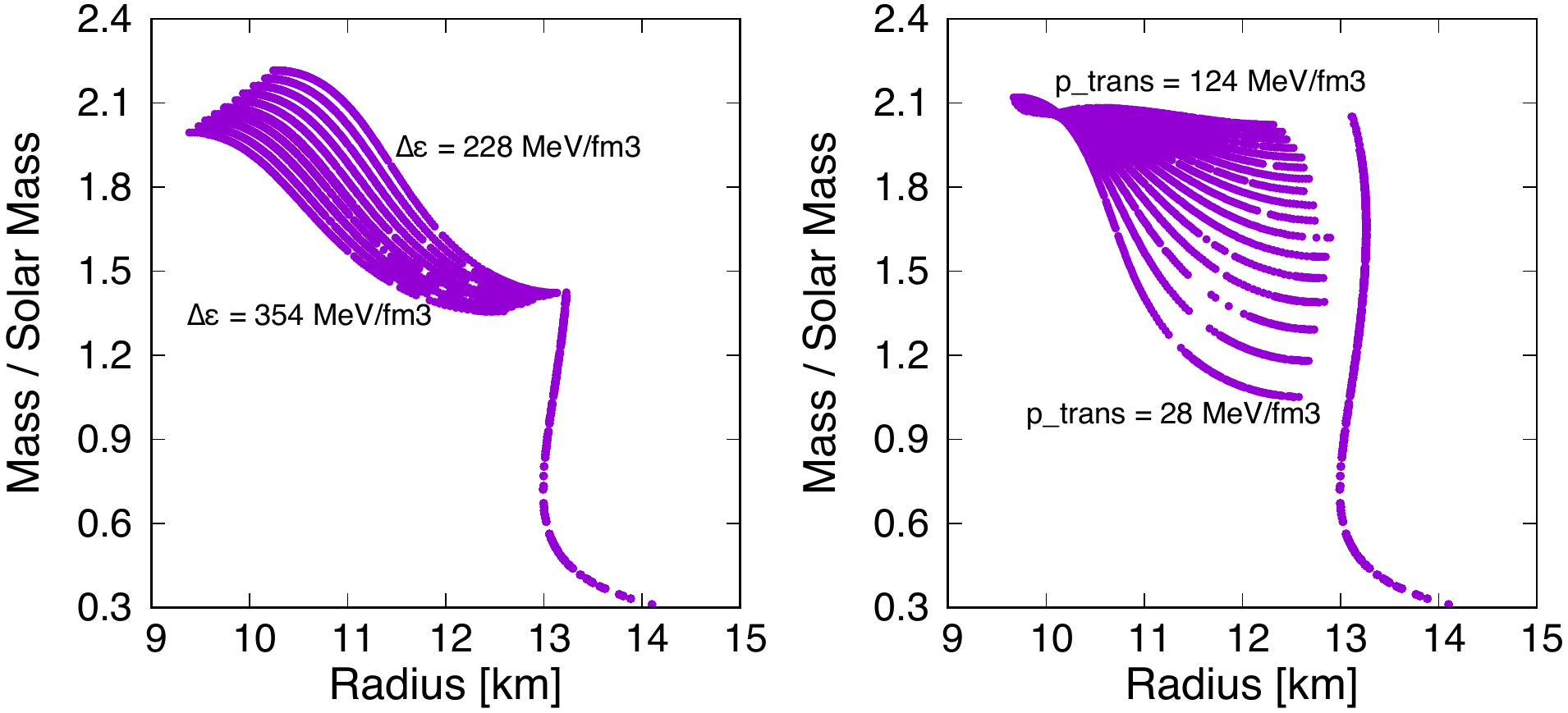}
  	\caption[Steepness]{\footnotesize In these diagrams multiple mass-radius relations with a constant $p_{trans}$ (left) and a constant $\Delta\epsilon$ (right) are depicted, by varying the other parameter. A change of $\Delta\epsilon$ results in a different position of the second maximum while the shape of the second branch remains nearly unaffected. The contrary is true for the a change in $p_{trans}$. The location of the second maximum remains nearly identical for different transition pressures while the shape of the second branch becomes much steeper for lower $p_{trans}$.}
  	\label{Steepness}
  \end{figure}

  The effects of varying $p_{trans}$ and $\Delta\epsilon$ on the mass-radius relation are seen in figure \ref{Steepness}. The shape of the second branch appears to be nearly unaffected by changes in $\Delta\epsilon$, see the left-hand side of figure \ref{Steepness}. The variation of $p_{trans}$ results in different slopes in the second branch with low values resulting in steeper curves. However, the position of the second maximum remains nearly constant for varying $p_{trans}$. We conclude that $\Delta\epsilon$ sets the maximum mass of the second branch, while $p_{trans}$ controls the slope of the mass radius relation of the second branch.
  
\begin{figure}
	\centering
	\includegraphics[width=8.5cm]{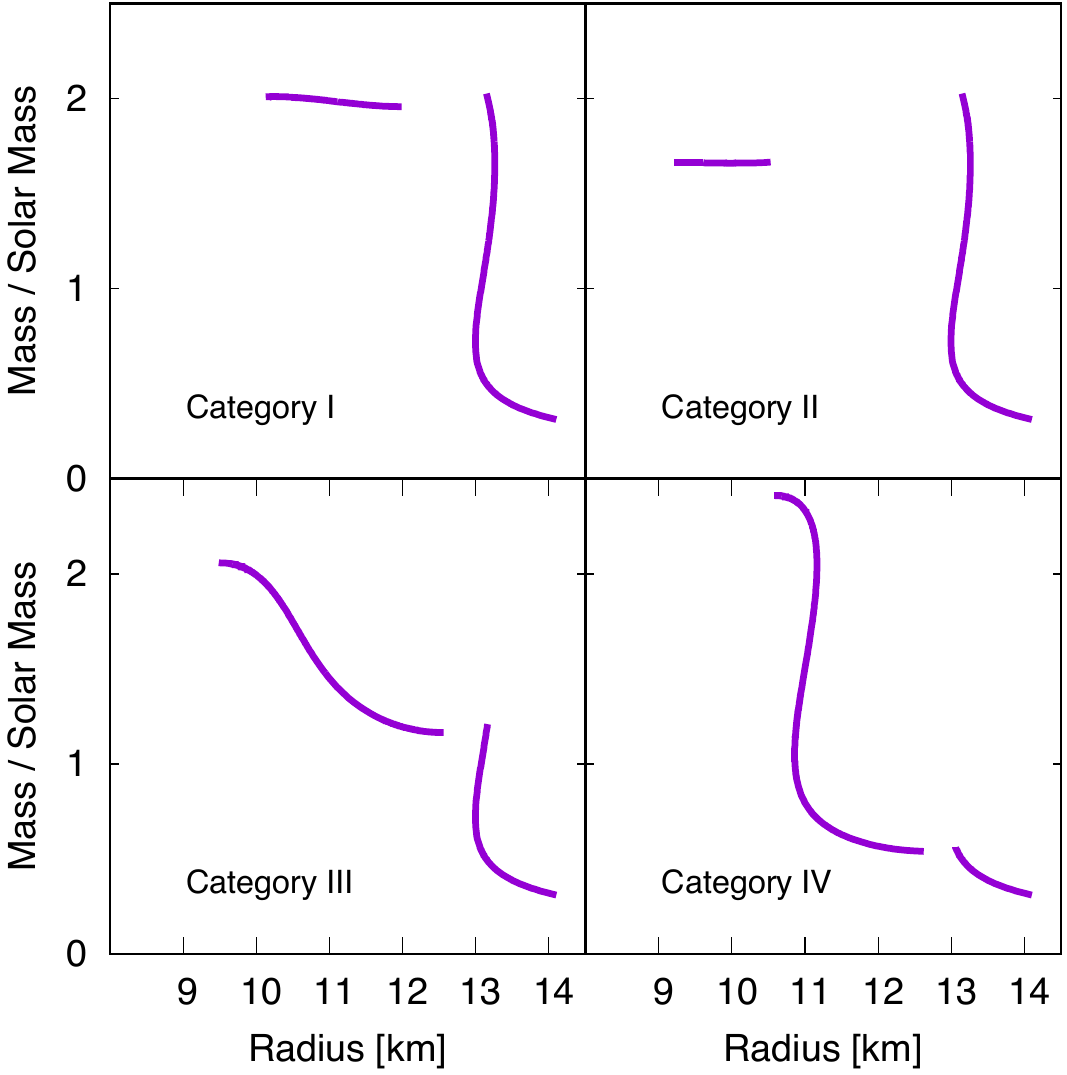}
	\caption[Classification]{\footnotesize Representative examples of the defined categories. Category \rom{1} and \rom{2} exhibit a flat second branch while the second branch of categories \rom{3} and \rom{4} are steeper.}
	\label{Multiplotclass}
\end{figure}

  Another important observation is that the value of $\Delta\epsilon$ has virtually no influence on the mass at the first maximum. This is due to the first branch becoming unstable at about the transitional pressure meaning that only the second branch is effected by $\Delta\epsilon$. Even though there are hybrid stars to be found in the first branch if the combinations of $\Delta\epsilon$ and $p_{trans}$ are below the Seidov-limit these stars have a negligible effect on the value of the first maximum \cite{Lindblom:1998dp}.\\
  Using this feature it is possible to assign a specific mass at the first maximum to a distinct $p_{trans}$. Likewise a relation between $\Delta\epsilon$ and the second maximum can be observed even though it is not as visible. With these relations it becomes possible to define four distinct categories in which the twin star solutions can be organized. Examples of these categories are shown in figure \ref{Multiplotclass} and defined as follows:
  
  \newpage
  		\begin{description}
  	\item[Category \rom{1}] defined by demanding that $M_{max_1} \ge 2M_\odot$ and $M_{max_2} \ge 2M_\odot$. That condition is provided by equations of state with $214\mathrm{MeV/fm^3} \lesssim \Delta\epsilon \lesssim 375\mathrm{MeV/fm^3}$ and $118\mathrm{MeV/fm^3} \lesssim p_{trans} \lesssim 184\mathrm{MeV/fm^3}$. In this category the second branch is nearly flat with $M_{min} \ge 2M_\odot$. This category features the heaviest twin stars with $M=2.24M_\odot$ for $p_{trans}=184\mathrm{MeV/fm^3}$ and $\Delta\epsilon=214\mathrm{MeV/fm^3}$ where stars of this mass are located in both branches. Apart from category \rom{4} this is the highest calculated mass.  
  \end{description}
  \begin{description}	
  	\item[Category \rom{2}] defined by demanding that the conditions $M_{max_1} \ge 2M_\odot$ and $M_{max_2} < 2M_\odot$ are satisfied. The equations of state for this category are provided by choosing values of $375\mathrm{MeV/fm^3} \lesssim \Delta\epsilon \lesssim 725\mathrm{ MeV/fm^3}$ and $118\mathrm{MeV/fm^3} \lesssim p_{trans} \lesssim 136\mathrm{MeV/fm^3}$.
  	 Akin to category \rom{1} the second branch has a nearly constant mass but with lower values. 
  \end{description}
  \begin{description}		 
  	\item[Category \rom{3}] defined by demanding that $M_{max_2} \ge 2M_\odot$ and $2M_\odot \ge M_{max_1} \ge 1M_\odot$. Equations of state contained in this category are generated by the conditions $214\mathrm{MeV/fm^3} \lesssim \Delta\epsilon \lesssim 368\mathrm{MeV/fm^3}$ and $24\mathrm{MeV/fm^3} \lesssim p_{trans} \lesssim 117\mathrm{MeV/fm^3}$. The second branch of hybrid stars rises much steeper in mass than in the previous categories. For increasing $p_{trans}$ as a function of radius the second branch shows a shape similar to categories \rom{1} and \rom{2}. 
  \end{description}
  \begin{description}		
  	\item[Category \rom{4}] defined by demanding that $M_{max_2} \ge 2M_\odot$ and $M_{max_1} \le 1M_\odot$. The equations of state for this category are provided by the values $150\mathrm{MeV/fm^3} \lesssim \Delta\epsilon \lesssim 425 \mathrm{MeV/fm^3}$ and $7 \mathrm{MeV/fm^3} \lesssim p_{trans} \lesssim 23 \mathrm{MeV/fm^3}$. The second branch increases in mass for nearly constant radii, leading to high masses. Only this category contains twin stars with masses of $1M_\odot$ or lower. Neutron stars formed in core-collapse supernovae have a lower limit of about one solar mass \cite{zdunikhaenselschaeffer:1983pti}.
  	The highest calculated mass in this category is $M=2.65M_\odot$ at $p_{trans}=10\mathrm{MeV/fm^3}$ and $\Delta\epsilon=150\mathrm{MeV/fm^3}$. 
  \end{description}
 
 \begin{table}
	\centering
	\begin{tabular}{l|c|c|c|c}
		& Low $p_{trans}$ & High $p_{trans}$ & Low $\Delta\epsilon$ & High $\Delta\epsilon$\\
		\hline
		Category \rom{1} & 118 & 184 & 214 & 375\\
		\hline
		Category \rom{2} & 118  & 136 &375 & 725\\
		\hline
		Category \rom{3} & 24 & 117 & 214 & 368\\
		\hline
		Category \rom{4} & 7 & 23 & 150 & 425\\
	\end{tabular}
	\caption{\footnotesize The four categories of twin stars defined by the masses of their maxima. All entries are given in units of  $\rm{MeV/fm^3}$. "High" and "Low" describes the upper or lower limit of $p_{trans}$ and $\Delta\epsilon$ of the category.}
	\label{tab:table1}
\end{table} 

\begin{figure}
	\centering
	\includegraphics[width=8.5cm]{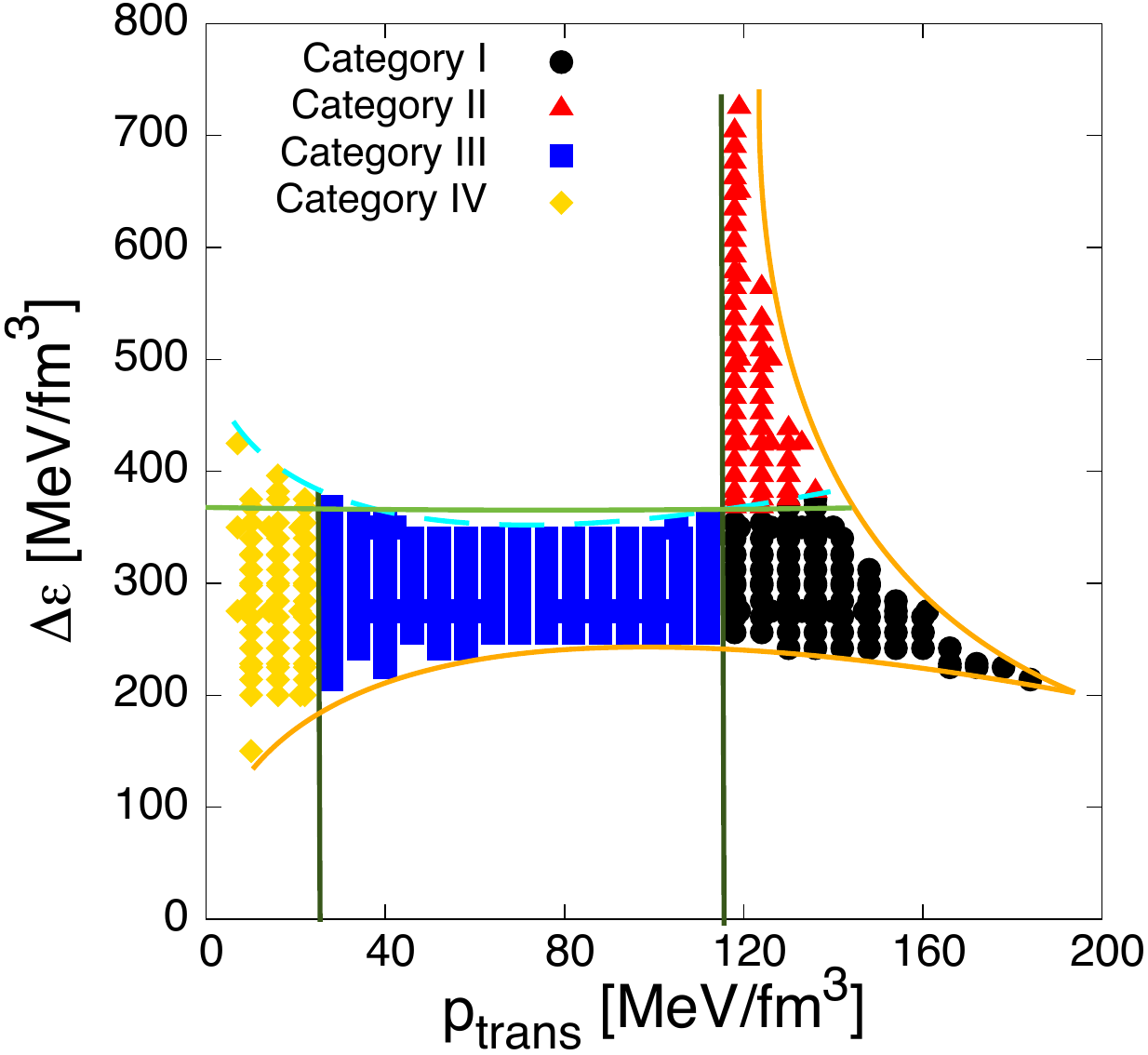}
	\caption[\footnotesize Allowed Area categorized]{Combinations of energy density difference $\Delta\epsilon$ and transitional pressure $p_{trans}$ leading to third families. The vertical continuous green lines indicate the $1M_\odot$ and the $2M_\odot$ limit at $M_{max_1}$ respectively. The turquois dashed line is the limit at which $M_{max_2}$ drops below $2M_\odot$. It can be roughly approximated with the continuous green line. Upon comparison with table \ref{tab:table1} one can see these lines correspond to the limits of the categories in the parameter space.}
	\label{allowedclass}
\end{figure}

The flat second branch is characteristic for category \rom{1} see the left upper graphic of figure \ref{Multiplotclass}. The mass of the first and second maximum are nearly identical in this category as well the mass at the minimum. In the upper right graphic an example of category \rom{2} is displayed. Like category \rom{1} the second branch is nearly flat, yet the masses at the maxima are significantly lower. In the left graphic of the second row a typical mass-radius relation of category \rom{3} is visible. The mass of the first maximum is located above $1M_\odot$ and the second branch is much steeper compared to previous categories. For category \rom{4} displayed in the right graphic of the second row, the hadronic branch becomes unstable at masses below $1M_\odot$. Here the second branch is even steeper than in category \rom{3}, showing a nearly constant radius for different masses.\\\\
In figure \ref{allowedclass} the twin star region with the four categories is depicted for different values of $p_{trans}$ and $\Delta\epsilon$. The turquois dashed line, separating category \rom{2} from all others, represents the maximal value of $\Delta\epsilon$ that still provides a mass of $2M_\odot$ at the second maximum. The relation between $p_{trans}$ and $M_{max_1}$ is indicated by the dark continuous vertical green line for the masses $1M_\odot$ and $2M_\odot$. The horizontal continuous green line indicates the approximate relation between $\Delta\epsilon$ and $M_{max_2}$ for $M_{max_2}=2M_\odot$.  

  \subsection{Radius Examination}

 The radius difference of twin star solutions is strongly influenced by the choice of $p_{trans}$ and $\Delta\epsilon$ due to their effect on position and shape of the second branch in the mass-radius diagram. We define the radius difference in the following for bins of $0.1M_\odot$.\\
 High values of $\Delta\epsilon$ lead to small radii which in combination with a high value of $p_{trans}$, can lead to radius differences of up to 4km for stars with identical masses.\\
Since nearly all twin stars with $M \ge 2.2M_\odot$ are part of category \rom{1}, radius differences for that mass range can only be defined for those masses. Radii differences are more than 2km in the parameter space of $\mathrm{260MeV/fm^3}\lesssim\Delta\epsilon\lesssim\mathrm{295MeV/fm^3}$ and $\mathrm{135MeV/fm^3}\lesssim p_{trans}\lesssim\mathrm{155MeV/fm^3}$. Category \rom{3} includes twin stars with radius differences nearly reaching 1km for $M \ge 2.2 M_\odot$. This is possible due to the comparatively small value of $\Delta\epsilon$ which allows for higher masses at $M_{max_2}$ in combination with large values of $p_{trans}$ a flat mass-radius curves for the second branch. All such twin stars in category \rom{3} are part of the second branch as the first branch never reaches the mass of $2M_\odot$ by definition.\\ 
For 2$M_\odot$ the point where category \rom{1}, \rom{2} and \rom{3} intersect reaches radius differences in excess of 3km. 
Twin stars with masses of $1.8M_\odot$ can be found in category \rom{2} and \rom{3}. Due to the mass-radius curves becoming steeper with decreasing $p_{trans}$ there are few cases within the third category that achieve a radius difference of $\Delta R=\mathrm{2km}$. Since the second category contains EoSs with higher transitional pressures the second branch is flat, resulting in radius differences of 3km for twin stars with $1.8M_\odot$ .\\
The radius differences for twin stars with masses of approximately $1.6M_\odot$ contain the largest value of $\Delta R$. A value of 4.0km is found in category \rom{2} for the parameters $p_{trans}=\mathrm{118eV/fm^3}$ and $\Delta\epsilon=690\mathrm{MeV/fm^3}$. Only a small percentage of category \rom{2} contains twin stars with masses of $M=1.6M_\odot$ due to high $\Delta\epsilon$ required. However, all of these EoSs result in large $\Delta R$. While category \rom{1} does not contain any twin stars with this mass category \rom{3} contains a small parameter space with $\Delta R > 1\mathrm{km}$.\\
Only a small portion of parameter space of category \rom{3} contains twin stars with masses of $1.4M_\odot$ and $\Delta R > 1\mathrm{km}$.\\
The same statement holds for less massive twin stars as the mass-radius diagrams become so steep that even radius differences of 1km are rare.\\
Table \ref{tab:table2} summarizes the largest radius differences for the corresponding twin star masses in each category. Category \rom{4} is not included in this table, as by definition no twin stars of over $1M_\odot$ are part of this category and radius differences in that category are very small.\\

\begin{figure}[H]
	\centering
	\includegraphics[width=8cm]{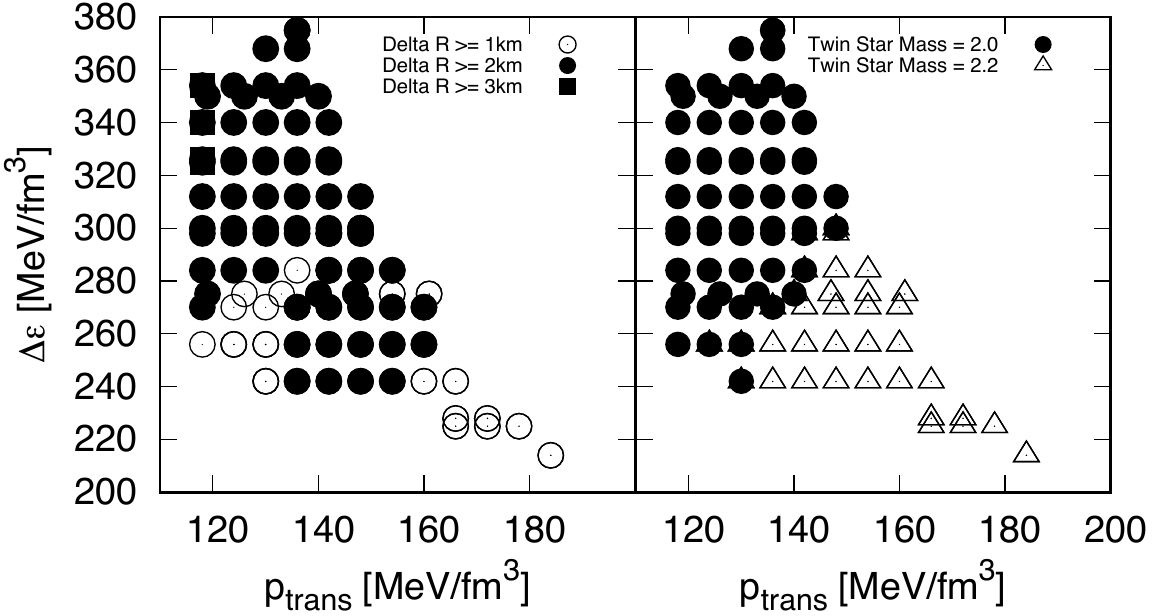}
	\caption[Summary]{\footnotesize Left: radius differences for twin stars in dependence on $p_{trans}$ and $\Delta\epsilon$ in category \rom{1}. Right:  most massive twin star configurations. The largest $\Delta R$ can be found for twin stars with masses of $M=2M_\odot$, specifically for $p_{trans}$ near the lower limit and $\Delta\epsilon$ near the upper limit of this category.} 
	\label{Endeplot1}
\end{figure}

\begin{figure}[H]
	\centering
	\includegraphics[width=8cm]{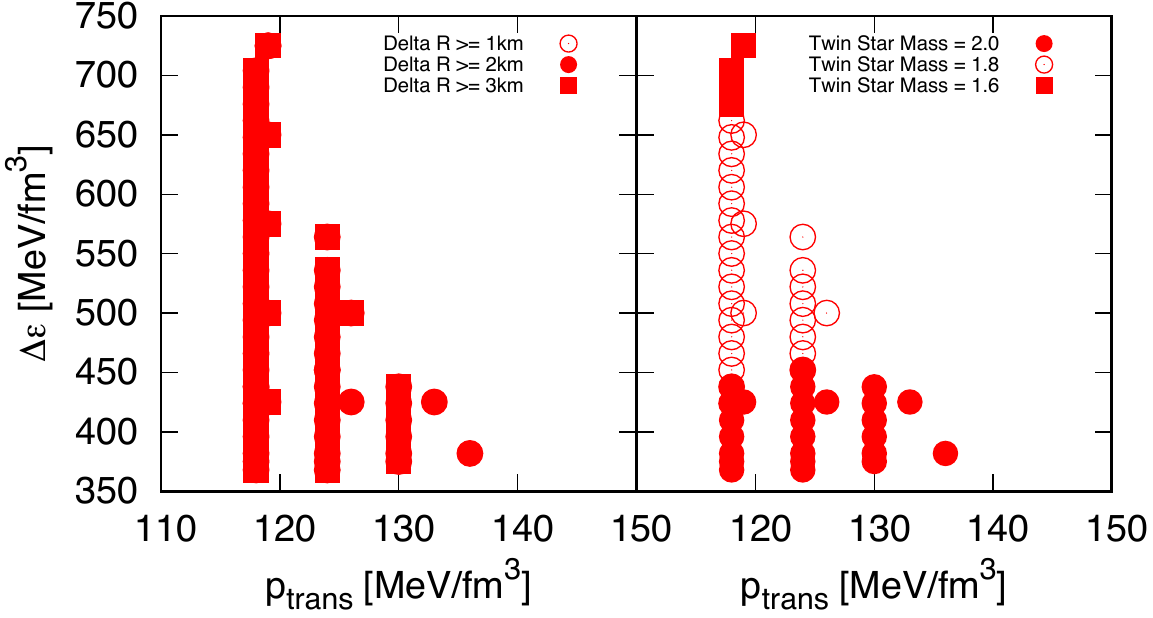}
	\caption[Summary]{\footnotesize Twin star masses (right) and the corresponding radius differences $\Delta R$ (left) of category \rom{2}. As in category \rom{1} the largest values of $\Delta R$ are located at the upper limit of $\Delta\epsilon$ and the lower limit of $p_{trans}$. However the corresponding twin star masses are lower.}
	\label{Endeplot2}
\end{figure}

\begin{figure}[H]
	\centering
	\includegraphics[width=8cm]{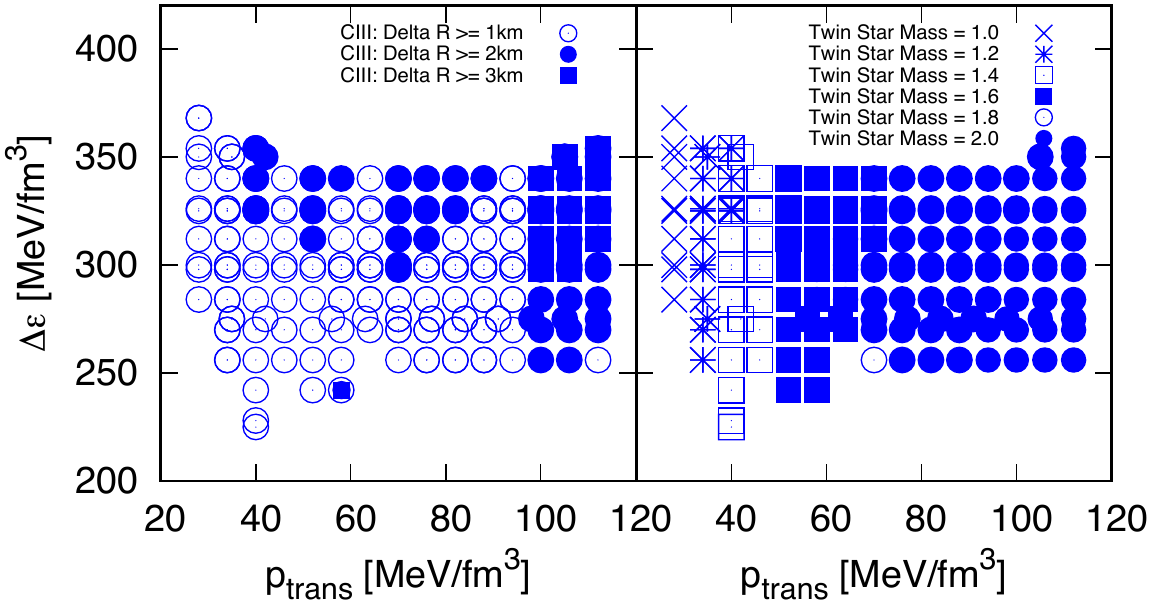}
	\caption[Summary]{\footnotesize Twin star masses (right) and the corresponding radius differences $\Delta R$ (left) of category \rom{3}. Contrary to previously discussed categories the largest radius differences are generated by $p_{trans}$ near the upper limit. This is due to the fact that on one hand larger values of $p_{trans}$ lead to flatter hybrid branches but on the other hand larger values of $p_{trans}$ move the second branch closer to the first one. As in the other categories large values of $\Delta\epsilon$ are beneficial for large radius differences and thus the largest values of $\Delta R$ in category \rom{3} are located near the upper limit of $\Delta\epsilon$.}
	\label{Endeplot3}
\end{figure}

In figures \ref{Endeplot1}-\ref{Endeplot3} the dependence of twin star mass and radius difference on $p_{trans}$ and $\Delta\epsilon$ is shown. On the left hand side of these figures the radius difference as a function of our parameters is depicted while on the right hand side the corresponding twin star mass as a function of $p_{trans}$ and $\Delta\epsilon$ is depicted. Figure \ref{Endeplot1} shows category \rom{1}. In this graphic the highest values of twin star mass are to be found as high values of $p_{trans}$ benefit high twin star masses.\\
The second category is shown in figure \ref{Endeplot2}. It contains the largest radius difference of about 4km. This can be attributed to the high values of $\Delta\epsilon$ that benefit large values of $\Delta R$.\\
Figure \ref{Endeplot3} shows the radius differences and twin star masses of category \rom{3}. The trend set by the previous categories that low values of $p_{trans}$ and high values of $\Delta\epsilon$ result in large radius differences is not continued in this category. This behavior is caused by low values of $p_{trans}$ resulting in a steeper hybrid star branch, with less massive twin stars with smaller radius differences.

  \begin{table}[h!]
	\centering
	\begin{tabular}{l|c|c|c|c|c}
		& $2.2M_\odot$  & $2.0M_\odot$ & $1.8M_\odot$  & $1.6M_\odot$ & $1.4M_\odot$ \\
		\hline
		Category \rom{1} & 2.41km & 3.09km & / & / & / \\
		\hline
		Category \rom{2} & / & 3.27km & 3.88km & 3.96km & /\\
		\hline
		Category \rom{3} & / & 3.18km & 2.63km & 2.27km & 2.25km\\
	\end{tabular}
	\caption{\footnotesize Largest radius differences for twin stars for the different categories \rom{1}-\rom{3}.The radius differences found in category \rom{4} only apply to stars with masses below $1M_\odot$.}
	\label{tab:table2}
\end{table}

  \section{Conclusions}\label{conclusion}
  In this work we employed a density dependent EoS, taken from \cite{Typel:2009sy}, for the nuclear shell, and a constant speed of sound parametrized EoS (CSS) for the core of a compact star. We study twin star solutions and their properties resulting from a first order phase transition between these two EoSs. Because of the 2M$_{\odot}$ constraint, stiff EoS are advantageous, so that our choice of the speed of sound within quark matter was assumed to be $c_s^2=1$.
 The two parameters $p_{trans}$ and $\Delta\epsilon$, describing the pressure at the transition and the accompanying jump in energy density, have thus the highest impact on the mass-radius relation and hence on the twin star solutions examined.\\
 We demonstrate that stable compact twin stars only exist within a small parameter space, which is narrowed down further by excluding all equations of state that do not reach 2$M_\odot$. 
 Since the first maximum in the mass-radius relation of twin stars is independent of $\Delta\epsilon$, and the second maximum on the other hand independent of $p_{trans}$ (see Fig.\ref{Steepness}) four distinct categories of twin stars can be defined within the investigated parameters $p_{trans}$ and $\Delta\epsilon$.\\
 Category \rom{1} contains mass-radius relations with masses of over $2M_\odot$ at both maxima and is located at values of $118 \mathrm{MeV/fm^3} \leq p_{trans}\leq 184 \mathrm{MeV/fm^3}$ and $214 \mathrm{MeV/fm^3} \leq \Delta\epsilon \leq 375 \mathrm{MeV/fm^3}$. The mass at the first maximum results from the transition pressure $p_{trans}$, where  $p_{trans}=\rm{118MeV/fm^3}$ corresponds to approximately $2M_{\odot}$.\\ 
 Category \rom{2} contains mass-radius relations with masses of over $2M_\odot$ only at their first maximum and is located at values of $118 \mathrm{MeV/fm^3} \leq p_{trans}\leq 136 \mathrm{MeV/fm^3}$ and $375 \mathrm{MeV/fm^3} \leq \Delta\epsilon \leq 725 \mathrm{MeV/fm^3}$. In this category the largest radius differences for twin stars can be found.\\ 
 Category \rom{3} contains mass-radius relations with $2M_\odot$ only at the second maximum but still more than $1M_\odot$ at the first maximum and can be found for values of $24 \mathrm{MeV/fm^3} \leq p_{trans}\leq 117 \mathrm{MeV/fm^3}$ and $214 \mathrm{MeV/fm^3} \leq \Delta\epsilon \leq 368 \mathrm{MeV/fm^3}$.\\ 
 Category \rom{4} contains mass-radius relations with masses of over $2M_\odot$ at the second maximum and masses under $1M_\odot$ at the first maximum. It can be found for values of $7 \mathrm{MeV/fm^3} \leq p_{trans}\leq 23 \mathrm{MeV/fm^3}$ and $150 \mathrm{MeV/fm^3} \leq \Delta\epsilon \leq 425 \mathrm{MeV/fm^3}$.\\ 
  
  Twin star solutions in general are not easy to model but have been found in many different kind of 
  phase-transition scenarios, e.g. hadron-quark phase transition  
  \cite{Mishustin:2002xe,Bhattacharyya:2004fn,Zacchi:2015oma}, hyperon phase transition \cite{SchaffnerBielich:2002ki}, pion- \cite{Kampfer:1981yr} and kaon condensation \cite{Banik04,Banik:2004fa}.\\ 
  In \cite{Zacchi:2015oma} for instance the combination of the DD2-EoS for the description of hadronic matter and a SU(3) chiral quark matter EoS for the stars core was examined. Twin star solutions could not be found due to the abrupt decrease of the  speed of sound $c_s^2$ at the phase transition contrary to the abrupt increase modeled here. It is interesting to note that in \cite{Zacchi:2016tjw} twin star solutions resulting from a SU(3) quark matter EoS only, with a chiral-like crossover phase transition, were found.
  Heiniman et. al. \cite{Heinimann:2016zbx} did a parameter scan with an identical parametrization to the one used in our approach, but with $c_s^2=1/3$. This leads to a much smaller area in parameter space containing twin star solutions with an even smaller parameter space containing stars with $M>2M_\odot$, which is in accordance with \cite{Alford:2013aca,Alford:2015dpa}.\\
  Blaschke et. al. \cite{Blaschke:2015uva} also found twin star solutions by assuming a first order phase transition.  
  Their twin star solutions are located near $2M_\odot$ and have very similar mass-radius relations to the ones found in category \rom{1} and \rom{3}.\\ 
   Alford et. al. \cite{Alford:2013aca,Alford:2015dpa} find third family solutions in a very similar parameter space for $c_s^2=1$.  
 They study the radius differences between hybrid stars but do not classify twin star solutions into different categories. However, the corresponding radius differences between twin stars have so far not been analyzed. Alford et. al. examine the radius differences between the maxima of the branches rather than the radius differences between stars with equal mass. This yields similar values of $\Delta R$ for categories \rom{1} and \rom{2}, as the maxima have approximately the same mass. However they found much larger values of $\Delta R$ in category \rom{3} due to their different approach.
  We find that comparing twin star masses with their corresponding radius differences $\Delta R$ would enable a unambiguous allocation to a small parameter space. These radius differences have values of up to $\Delta R=4$km in category \rom{2} and are larger than 1km for the majority of all other categories as well.\\
  The space mission NICER (Neutron star Interior Composition ExploreR) \cite{2014SPIE.9144E..20A} will be able to measure masses and radii of compact objects with a resolution in radius of $\Delta R\sim 1$~km. The discovery of two compact objects with same masses but different radii could be indeed signal the existence of twin stars, which implies the existence of a phase transition in dense matter \cite{Zacchi:2016tjw}.

\begin{acknowledgements}
AZ is supported by the Helmholtz Graduate School
for Heavy-Ion Research (HGS-HIRe), the Helmholtz Research 
School for Quark Matter (H-QM) and the Stiftung Giersch.
\end{acknowledgements}
\bibliographystyle{apsrev4-1}
\bibliography{biblio}
\end{document}